\documentclass[11pt]{article}

\usepackage{epsfig}
\usepackage{latexsym}
\usepackage{amssymb}

\newcommand{\sect}[1]{\setcounter{equation}{0}\section{#1}}

\newcommand{\reef}[1]{(\ref{#1})}
\def\be{\begin{equation}}
\def\ee{\end{equation}}
\def\ba{\begin{eqnarray}}
\def\ea{\end{eqnarray}}
\def\prt{\partial}

\begin{document}

\thispagestyle{empty}
\rightline{\small hep-th/0209141 \hfill  RUNHETC-2002-27} 
\vspace*{2cm}

\begin{center}
{ \LARGE {\bf Moduli and Brane Intersections}}\\[.25em]
\vspace*{1cm}

Neil D. Lambert\footnote{nlambert@physics.rutgers.edu}
\\
\vspace*{0.2cm}
{\it Department of Physics and Astronomy}\\
{\it Rutgers University}\\  
{\it Piscataway NJ 08855}\\ 
{\it USA}\\

\vspace{2cm} ABSTRACT
\end{center}
\noindent
We discuss the worldvolume description of intersecting D-branes, including
the metric on the moduli space of deformations.  
We impose a choice of static gauge that treats 
all the branes on an equal footing and 
describes the intersection of 
D-branes as an embedded special Lagrangian three-surface. 
Some explicit solutions to
these equations are given  and their interpretation
in terms of a superpotential on moduli space is discussed.
These surfaces are interpreted in terms of the 
flat direction of a non-Abelian superpotential
and imply the existance of non-compact $G_2$ manifolds.

\vfill \setcounter{page}{0} \setcounter{footnote}{0}
\newpage

\sect{Introduction}

D-branes provide a fascinating bridge between the  classical geometry of
minimal surfaces and  quantum  non-Abelian gauge theory.
The usual set-up consists of several branes 
intersecting so that their common worldvolume has some
Poincare invariance and preserves some supersymmetry. From the microscopic
D-brane perspective the low energy dynamics is governed by open string
theory and hence coincides with Yang-Mills theory to lowest order in 
derivatives. In the macroscopic
description branes are minimal volume surfaces their dynamics is essentially
determined by geometry. To be more precise the induced metric on the brane
defines a metric on the moduli space of deformations of the surface and the
low energy effective action is then given by the associated
$\sigma$-model. 

Recently,
in connection with 
compactifications of M-theory on exceptional manifolds, brane intersections
with minimal supersymmetry have been studied and related to gauge theory
\cite{aw,gt,gst}.
Supersymmetry imposes the additional condition that the brane worldvolume is
not just minimal but calibrated \cite{BBS}.
In \cite{GLW} 
Bogomoln'yi equations  were derived for various intersections where static
gauge was imposed with respect to one of the branes. The resulting equations
are known as Harvey-Lawson equations \cite{HL} 
and are also known to be local conditions
for the embedded surface to be calibrated. Here we wish to revisit some of this
analysis but treat all the branes on an equal footing. In particular 
we will not
impose static gauge for any one particular brane. A  benefit of this
analysis is that near the intersection of two branes the scalar fields
behave smoothly (although they will now diverge linearly at
infinity). 

This paper is organized as follows. In section 2 we
discuss some general features of supersymmetric embeddings of D-branes
into spacetime. In particular we obtain a general form for the metric
on the moduli space of deformations which yields a simple condition for
the normalisability of the moduli.  In section 3 we specialise to
the case of three-dimensional intersections.
Our discussion will be in terms of D3-branes but can readily be applied to
other cases, most notably D6-branes. In particular we discuss solutions with
preserve eight and four supercharges and correspond to Riemann surfaces and
special Lagrangian surfaces respectively. 
We also include a 
discussion of the moduli spaces and the relation to 
flat directions of a non-Abelian superpotential.
Finally in section 4 
we consider the analogous case of intersecting D6-branes and
their lift to new $G_2$ manifolds in M-theory.

\sect{Supersymmetric Embeddings}

We consider D$p$-branes where the worldvolume gauge field vanishes
and the background spacetime is flat.  Most of our
analysis can also be readily extended to cases where the background is
of special holonomy, although concrete results will be considerably more
difficult to obtain.
The condition that the brane is supersymmetric is 
$\Gamma\epsilon_1=\epsilon_2$ where $\epsilon_1$ and $\epsilon_2$ are
the ten-dimensional supersymmetry generators. 
The projector $\Gamma$ is \cite{bt}
\be
\Gamma = {1\over (1+p)!}{1\over \sqrt{-{\rm det}(g)}}\epsilon^{i_0...i_p}
\prt_{i_0} X^{m_0}...\prt_{i_p} X^{m_p}\Gamma_{m_0...m_p}\ .
\label{Gamma}
\ee
Here $\Gamma_m$, $m=0,...,9$ form the ten-dimensional Clifford algebra.
The indices $i,j...=0,1,...,p$ label the worldvolume coordinates $\sigma^i$ of
the brane and $g_{ij} = \prt_iX^m\prt_jX^n \eta_{mn}$ is the induced metric
on the brane. In this paper we will assume that all branes are static
and fix $\sigma^0=X^0$.  Thus what follows $i,j=1,...,p$.

The important constraint of this system is that $\epsilon_1$ and $\epsilon_2$
are constant spinors. Thus finding a supersymmetric brane configuration
requires finding an embedding such that $\Gamma\epsilon_1$ is constant over
the brane. 
In particular, since $\Gamma^2 = \pm1$ and $\Gamma^\dag = \pm\Gamma$,
where the sign depends on $p$, we can follow \cite{BBS} and observe that
\be
\epsilon_2^\dag\epsilon_2+\epsilon_1^\dag\epsilon_1
\mp\epsilon_1^\dag\Gamma\epsilon_2
- \epsilon_2^\dag\Gamma\epsilon_1
=(\epsilon_2- \Gamma\epsilon_1)^\dag(\epsilon_2- \Gamma\epsilon_1)
\ge 0\ ,
\label{ineq}
\ee
with equality iff the embedding is supersymmetric, {\it i.e.} 
$\epsilon_2=\Gamma\epsilon_1$. If we normalize
$\epsilon_2^\dag\epsilon_2=\epsilon_1^\dag\epsilon_1=1$ and multiply by 
$\sqrt{-{\rm det}(g)}d\sigma^1\wedge ...\wedge d\sigma^p$ we find the
inequality
\be
\sqrt{-{\rm det}(g)}d\sigma^1\wedge ...\wedge d\sigma^p \ge 
\omega_{m_1...m_p}
\prt_{1} X^{m_1}...\prt_{p} X^{m_p}
d\sigma^1\wedge ...\wedge d\sigma^p\ ,
\label{bound}
\ee
where
\be
\omega_{m_1...m_p}=
{1\over 2}\epsilon_2^\dag\Gamma_{0m_1...m_p}\epsilon_1
+c.c\ .
\label{cal}
\ee
The inequality \reef{bound} states that the volume form on the 
brane's worldvolume, when evaluated on any tangent plane,  
is bounded below by the pull-back to the
worldvolume of the closed spacetime $p$-form $\omega$.
Note that the same argument applies to ${\bar D}p$-branes where 
$\Gamma\epsilon_1=-\epsilon_2$
but as a consequence the sign of $\omega$ is reversed in \reef{bound}. 

This is precisely the notion of a calibration {\it i.e.} a closed $p$-form
$\omega$ in the bulk spacetime whose integral over $p$-dimensional
surfaces is bounded above by their volume.
A surface which  saturates the bound, {\it i.e.} a supersymmetric embedding,  
is said to be calibrated by  $\omega$.
It is straightforward to show that any smooth 
calibrated surface is volume minimizing
in its homology class. To see this we suppose that a surface
$\Sigma$ is calibrated and consider a perturbation to a new manifold
$\Sigma'$:\footnote{Since $\Sigma$ will generically be non-compact we require
that $\Sigma'$ is different from $\Sigma$ only on a compact set
$\Sigma_c$  and that all the integrals are over $\Sigma_c$}.
\ba
{\rm Vol}(\Sigma) = \int_\Sigma \omega 
= \int_{\partial B} \omega +\int_{\Sigma'}\omega 
= \int_{\Sigma'} \omega \le {\rm Vol}(\Sigma')\ .
\ea
Here $B$ is a compact manifold whose boundary is $\Sigma \cup \Sigma'$
and we have used the fact that
\be
\int_{\Sigma}\omega - \int_{\Sigma'}\omega 
=\int_{\partial B}\omega = \int_B d\omega = 0 \ .
\ee
Thus a calibrated surface satisfies the equation of motion for
a D$p$-brane (in the absence of other worldvolume fields).

Given a smooth embedding of a brane into space it will generally depend on
several moduli labeled by $u^a$, $a=1,...,n$. It is natural to consider
the low energy effective action of the brane, where its
moduli become time-dependent. More generally if 
the brane has an additional isometric 
direction, such as two D3-branes intersecting over a line, then one
can also introduce dependence on these extra spatial dimensions. 
The result will then be an effective field theory which is a trivial
extension of the action will now derive. In static gauge the
induced metric is
\be
g_{ij} = \partial_i X^m \partial_j X^n \eta_{mn} = 
\eta_{ij} + \partial_i X^I\partial_j X^J \delta_{IJ}\ ,
\label{induced}
\ee
where $\sigma^i=X^i$ and $I=p+1,...,9$
labels the transverse coordinates. The (potential) 
energy for a static D$p$-brane  simply its spatial volume
\be
E = \int d^{p}\sigma \sqrt{-{\rm det}(g)}\ .
\label{energy}
\ee
To obtain an effective Lagrangian for the moduli $u^a$ we allow them to
pick up a slow time dependence so that the induced metric \reef{induced}
also has $g_{00}$ and $g_{0i}$ components. The effective Lagrangian 
${\cal L}_{eff}$ is then just  (minus) the potential energy functional 
$E$ expanded
to second order in time derivatives but to all orders in
the spatial derivatives.
If we use a dot to denote a time derivative
then we find that expanding \reef{energy} to second order gives
\be
{\cal L}_{eff} = \gamma_{ab}\dot u^a\dot u^b\ ,
\label{modact}
\ee
where
\be\label{modmetric}
\gamma_{ab}=\int d^p\sigma \sqrt{{\rm det}(1+MM^t)} 
{\partial X^I\over \partial u^a }\left({1\over 1 + M^tM}\right)_{IJ}
{\partial X^J\over \partial u^b }\ ,\quad M_i^{\ J} = \partial_i X^J\ ,
\ee 
and we have dropped the constant but  infinite  
volume of a static D$p$-brane. 
In deriving \reef{modmetric} we have used the matrix identity 
$1-N^t(1+NN^t)^{-1}N = (1+N^tN)^{-1}$. 
Note also that
for any real  matrix $N$, $N^tN$ is  symmetric with non-negative
eigenvalues. Hence $1+MM^t$ and $1+M^tM$ 
are always positive definite and therefore 
the inverse of  the matrix $1+M^tM$  
exists everywhere and is positive semi-definite
(it can vanish at points were $M^tM$ diverges).

For some moduli the integral in \reef{modmetric}
will not converge and these moduli must
become frozen at low energy. However for other modes, 
{\it i.e.} those whose  
effect on the embedding vanishes sufficiently quickly at infinity, 
the integral converges so that the action will have a finite kinetic term. 
Such moduli are said
to be normalisable and are dynamical at low energy. 
The effective action is then a
$\sigma$-model on the space of normalisable deformations of the surface
with the  metric  \reef{modmetric}.

In this paper we do not gauge fix to any particular brane. An
advantage of this choice is that the solutions are smooth embeddings at
at all finite values of $\sigma^i$ and asymptotically 
approach flat branes for large $\sigma^i$. 
Therefore the only divergences in the integral
\reef{modmetric} arise from the limit of large $\sigma^i$ where $M$ becomes
a constant matrix. 
Hence a modulus $u^a$ is normalisable if 
$\partial X^I/\partial u^a$  falls off strictly 
faster than $|\sigma|^{-p/2}$. This is analogous to a similar condition
observed in \cite{aw} for normalisable metric moduli. 

\section{Three-Dimensional Intersections}

For the rest of this paper we will restrict our attention to  
the case of intersecting D3-branes which preserve at least four supercharges. 
Our discussion also applies to various other intersecting D$p$-branes.
For example we could replace all the D3-branes
by D6-branes which intersect over a $3+1$-dimensional flat space. 
The restriction to examples which preserve four supercharges 
guarantees that the wrapped branes enjoy the equivalent of four-dimensional
$N=1$ supersymmetry.
We will also discuss cases which preserve more supersymmetry
where two D3-branes intersect on a line, which can alternatively
replaced by two D2-branes intersecting over a point or two D5-branes
intersecting over a $3+1$-dimensional space.
 
We begin with the most general  configuration of perpendicularly intersecting
D3-branes which preserves four supercharges  
\ba
\matrix{
D3:&1&2&3&&&\cr
\bar{D}3:&1&&&4&5&\cr
\bar{D}3:&&2&&4&&6\cr 
\bar{D}3:&&&3&&5&6\cr
}
\label{figtwo}
\ea
Note that for intersecting D3-branes there is a more general configuration but
it only preserves two supercharges and corresponds to associative three-cycle
in ${\bf R}^7$, for example see \cite{GLW}. 
In particular the branes in \reef{figtwo} preserve the supersymmetries 
\ba
\Gamma_{0123}\epsilon_1 &=& \epsilon_2\ ,\nonumber\\
\Gamma_{0154}\epsilon_1 &=& -\epsilon_2\ ,\nonumber\\
\Gamma_{0624}\epsilon_1 &=& -\epsilon_2\ ,\nonumber\\
\Gamma_{0653}\epsilon_1 &=& -\epsilon_2\ ,\nonumber\\
\label{susy}
\ea
respectively. The product of any three of these projectors gives the
fourth. Hence 
given any three of the branes in \reef{figtwo} one
finds that the fourth can be included without breaking any additional
supersymmetries.  In addition these preserved supersymmetries also allow
for D3-branes which do not intersect at right angles.
In total $1/8$ of the spacetime supersymmetries are
persevered by all the branes. The choice of D3-brane or 
${\bar {\rm D}}$3-brane is
arbitrary for any three of these branes, however once this choice is made
there is no such freedom for the  fourth brane. 
We have made the above choice to simplify the calculations below.

If we introduce complex coordinates\footnote{We will have no need for 
the remaining three coordinates $X^7,X^8,X^9$ in this 
paper.}  $Z^1 = X^1+iX^6$, $Z^2=X^2+iX^5$ and 
$Z^3 = X^3+iX^4$
then we can construct the three-form 
$dZ^1\wedge dZ^2\wedge dZ^3$ of ${\bf C}^3 = {\bf R}^6$
and also the calibration
\ba
\omega &=& {\rm Re}(dZ^1\wedge dZ^2\wedge dZ^3)\nonumber\\
&=&dX^1\wedge dX^2\wedge dX^3-dX^1\wedge dX^5\wedge dX^4
-dX^6\wedge dX^2\wedge dX^4-dX^6\wedge dX^5\wedge dX^3\ .\nonumber\\
\label{caltwo}
\ea
Note that, after imposing the constraints \reef{susy} this expression for
$\omega_{mnp}$ 
agrees with \reef{cal}.
Clearly $\omega$ calibrates each of the brane worldvolumes in \reef{figtwo}, 
{\it i.e.} the pull back of $\omega$ to the worldvolume is the volume form
(or minus the volume form for the anti-branes). 

Following \cite{taylor} we introduce new coordinates $Y^m$
\ba
Y^1 &=& {1\over\sqrt{2}}(X^1+X^6)\ , \quad Y^2 = {1\over\sqrt{2}}(X^2+X^5)
\ , \quad Y^3 = {1\over\sqrt{2}}(X^3+X^4)\ ,\nonumber\\
Y^4 &=& {1\over\sqrt{2}}(X^1-X^6)\ ,\quad  Y^5 = {1\over\sqrt{2}}(X^2-X^5)
\ ,\quad  Y^6 = {1\over\sqrt{2}}(X^3-X^4)\ .\nonumber\\
\label{coords}
\ea
The point of the $Y^m$ coordinates is that they do not favour any one
brane over the others.
Our next step is to impose the static gauge $\sigma^i = Y^i$, 
$i=1,2,3$. The coordinates $Y^I$, $I=4,5,6$ are now viewed as functions
of $\sigma^i$. 
In effect we have introduced an imaginary brane which lies
at the same angle with each of the branes in \reef{figtwo}.

Next we need to solve $\Gamma\epsilon_1=\epsilon_2$, where $\Gamma$ is
given in \reef{Gamma}, for constant spinors $\epsilon_1$ and $\epsilon_2$
which satisfy the constraints \reef{susy}.
It is helpful to introduce the matrix $M_i^{\ j} = \prt_iY^{j+3}$ and
also let ${\hat\Gamma}^i = \Gamma^{i+3}$ (so that $\Gamma_i$ and 
${\hat \Gamma}^i$ form two anti-commuting copies of the three-dimensional
Clifford algebra). The supersymmetry projections \reef{susy} are now
equivalent to
\ba
\Gamma^1{\hat \Gamma}_1\epsilon_1&=&\Gamma^2{\hat \Gamma}_2\epsilon_1=
\Gamma^3{\hat \Gamma}_3\epsilon_1 \nonumber\\
{1\over\sqrt{2}}\Gamma_{0}(\Gamma_1+\hat\Gamma_1)\Gamma_{23}\epsilon_1
&=&{1\over\sqrt{2}}\Gamma_{01}(\Gamma_2+\hat\Gamma_2)\Gamma_{3}\epsilon_1
={1\over\sqrt{2}}\Gamma_{012}(\Gamma_3+\hat\Gamma_3)\epsilon_1
=\epsilon_2
\ .\nonumber\\
\label{name}
\ea
Expanding out
$\Gamma\epsilon_1=\epsilon_2$ leads to 
\ba
\epsilon_2 &=& {\Gamma_{0123}\over \sqrt{-{\rm det}(g)}}\left\{
1+{1\over2}\sum_{i<j}(M_i^{\ j}-M_j^{\ i})\Gamma^i\hat\Gamma_j
+{\rm tr}(M)\Gamma^{1}\hat\Gamma_1 \right.\nonumber\\
&&\left.\ \ \ \ \ \ \ \ \ \ \ \ - {\rm det}(M)\Gamma^{1}\hat\Gamma_1
-{1\over2}\epsilon^{ijk}\epsilon_{lmn}\Gamma_k\hat\Gamma^n
\Gamma^{1}\hat\Gamma_1
M_i^{\ l}M_j^{\ m}
\right\}\epsilon_1\ .\nonumber\\
\label{expand}
\ea
Terms involving $\Gamma^i\hat\Gamma_j$ and 
$\Gamma^i\hat\Gamma_j\Gamma^1\hat\Gamma_1$with $i \ne j$ must vanish to
ensure that $\Gamma\epsilon_1=\epsilon_2$. The vanishing of the
$\Gamma^i\hat\Gamma_j$, $i \ne j$ terms  imply that 
\be
M_i^{\ j}=M_j^{\ i}\ .
\label{slagone}
\ee 
In addition \reef{slagone} also ensures the vanishing of
the $\Gamma^i\hat\Gamma_j\Gamma^1\hat\Gamma_1$, $i \ne j$ terms.
Once this is imposed the remaining
terms can be rearranged and the condition 
$\Gamma\epsilon_1= {1\over\sqrt{2}}\Gamma_{0}(\Gamma_1+\hat\Gamma_1)\Gamma_{23}\epsilon_1=\epsilon_2$ becomes
\be
{\rm tr}(1+M) = {1\over2}{\rm det}(1+M)+2
\ .
\label{slag}
\ee

In general the  
condition \reef{slagone} is solved by writing $Y^{j+3}=\prt_j F$ for some
potential $F(\sigma^i)$. The first condition then becomes a
second order  non-linear differential equation for $F$.
We note that $M=0$, {\it i.e.} a brane lying in the $Y^1,Y^2,Y^3$ plane,
is not supersymmetric (with respect to the supersymmetries \reef{susy}). 
With these conventions the branes in \reef{figtwo} correspond
to the solutions 
\ba 
M &=& {\rm diag}(+1,+1,+1)\ ,\nonumber\\
M &=& {\rm diag}(+1,-1,-1)\ ,\nonumber\\
M &=& {\rm diag}(-1,+1,-1)\ ,\nonumber\\
M &=& {\rm diag}(-1,-1,+1)\ ,\nonumber\\
\label{csol}
\ea 
respectively. 

We can also find the condition that $\Gamma\epsilon_1=  -\epsilon_2$.
From \reef{expand} we now find
\be
{\rm det}(1+M) = 2{\rm det}(M)+2\ ,\quad M_i^{\ j}=M_j^{\ i}\ .
\label{slagg}
\ee
Flipping all the signs in \reef{csol} gives solutions to
\reef{slagg} and describes the intersection
\ba
\matrix{
{\bar D}3:&&&&4&5&6\cr
{D3}:&&2&3& & &6\cr
{D3}:&1&&3&5&\cr 
{D3}:&1&2&&4&&\cr
}
\label{fignone}
\ea
However we will not devote any more time here to these embeddings.

Let us compare these embedding conditions 
with the equation that results if we had instead imposed
static gauge with respect to the first D3-brane, {\it i.e } 
$\sigma_1 = X^1,\sigma_2=X^2,\sigma_3=X^3$. The Bogomolnyi condition was
found in  \cite{GLW} to be 
\be\label{oldslag}
{\rm det}(\tilde M)  = {\rm tr}(\tilde M)\ ,\quad \tilde M_{i}^{\ j}=
\tilde M_j^{\ i}\ ,
\ee
where $\tilde M_i^{\ j} = \partial_iX^{j+3}$. This is the original Harvey-Lawson
equation  for a special Lagrangian embedding \cite{HL}.
While this equation is somewhat simpler than \reef{slag} or \reef{slagg}, 
it has the disadvantage that only the first brane 
in \reef{figtwo} appears as a solution; namely $\tilde M=0$.

Finally, if \reef{slag} is satisfied then
\be
\sqrt{-{\rm det}(g)} =
{2}|{\rm tr}(M)- {\rm det}(M))|\ ,
\label{detg}
\ee
which is a total derivative.
Note that the condition ${\rm det}(g) =0$ is the same as the 
the embedding equation \reef{oldslag} and corresponds to
a surface that is calibrated with respect to
$\tilde \omega={\rm Re}((dY^1+idY^6)\wedge (dY^2+idY^5)\wedge (dY^3+idY^4))$. 
However, as we saw above,
$1+MM^t$ is positive definite and therefore ${\rm det}(g)$
never vanishes. Hence surfaces which are calibrated
with respect to $\omega$ and $\tilde \omega$ have no common tangents.

\subsection{Two D3-branes intersecting over a line}

By describing intersecting branes as a single brane embedded into
spacetime we are implicitly assuming that the intersection point
has been smoothed-out so that a manifold structure exists. Thus the
general configuration of intersecting branes is realised as a smooth solution
$M(\sigma^i)$ to \reef{slag} which tends to the  constant solutions \reef{csol}
in various limits. That such solutions exist is less clear. 
Let us therefore start by discussing the 
simpler case of two intersecting D3-branes over a line. 
Here we can exploit
the underlying holomorphic structure to  construct all the necessary 
solutions.

If we remove the first two D3-branes from \reef{figtwo} 
we obtain the brane configuration
\ba
\matrix{
\bar{D3}:&2&&4&&6\cr
\bar{D3}:&&3&&5&6\cr 
}
\label{figone}
\ea
These branes preserve the supersymmetries
\be
\Gamma_{0624}\epsilon_1 =-\epsilon_2\ ,\quad 
\Gamma_{0653}\epsilon_1 = -\epsilon_2\ ,
\label{susytwo}
\ee
respectively. Hence they preserve a common 
$1/4$ of the spacetime supersymmetries.

From the worldvolume of our imaginary brane 
this is achieved by fixing $Y^4 =-\sigma^1$ and restricting  $Y^5$ and $Y^6$
to be independent of $\sigma^1$.  The Bogomoln'yi conditions \reef{slag} 
then reduce to 
\be
\prt_2Y^5=-\prt_3Y^6\ ,\quad \prt_2 Y^6=\prt_3 Y^5\ .
\label{Bogoone}
\ee
These equations are 
readily recognized as the Cauchy-Riemann equations
which are solved by introducing complex coordinates 
\be
Z = Y^6+iY^5\ ,\quad 
w = \sigma^2+i\sigma^3\ . \label{cxpcoords}
\ee
In this case the full non-linear 
Bogomoln'yi equations reduce to \cite{3brane} 
\be
\bar \prt Z=0\ .
\label{complex}
\ee

Solutions to \reef{complex} that represent intersecting branes 
are of course easy to constuct following \cite{witten}. 
They may be
implicitly described by a holomorphic equation $E(Z,w)=0$.
We are interested two intersecting branes so that  for each $w$ (except for
some isolated points) there should be two solutions for $Z$. Thus we
write $E = A(w)Z^2 + B(w)Z + C(w)$. Furthermore, for $w \rightarrow\infty$, the
branes should resemble flat planes, {\it i.e.} $Z \sim \lambda w$ in which
case the angle between the two branes is
$\theta = 2{\rm arctan}|\lambda|$. In addition there should be no other
points where $Z\rightarrow \infty$. Thus $A=1$, $B$ should be linear in
$w$ and $C$ quadratic. Without loss of generality we may shift $Z$ to
set $B=0$ and write $C(w) = \lambda^2(w-w_0)^2-\alpha^2$ so that
\be
Z = \pm\sqrt{\alpha^2-\lambda^2(w-w_0)^2}\ .
\label{Zsol}
\ee 
An artist's impression of part of this curve is in figure 1,  the full curve
is a two-dimensional surface in four dimensions and is connected. 
The solid line represents the curve and the dashed lines represent
the undeformed intersection.
The two branches, {\it i.e.} choices of
sign, correspond
to the top and bottom half of the curve.
There are
two branch 
points $w= w_0\pm \alpha/\lambda$ and so the full solution consists of
two copies of the complex $w$ plane.  Clearly $\alpha$ plays
the role of smoothing out the intersection,  $w_0$ describes the location
of the intersection and $\lambda$ controls the angle
between the two branes. 
As $w\rightarrow
\infty$ we find that \be
M \rightarrow \left(
\matrix{-1&0&0\cr
0&\pm{\rm Re}(\lambda)&\mp {\rm Im}(\lambda)\cr
0&\mp {\rm Im}(\lambda)&\mp {\rm Re}(\lambda)\cr}\right)
\ .
\label{Masy}
\ee
Thus for $\lambda^2 = 1$ these are indeed 
smooth curves that interpolate between
two of the constant solutions in \reef{csol}.

\begin {figure}
\center{ \epsfig{file=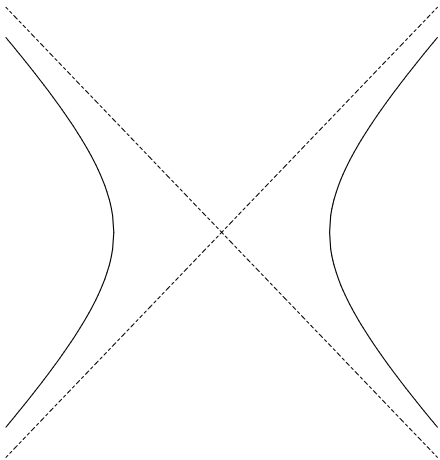,height=6cm,width=5.75cm}}
\caption{A holomorphic embedding of 2 D3-branes}
\end{figure}

We could also
try to find smooth solutions which interpolate between other pairs of 
branes in \reef{figtwo}. For the middle two branes we perform an analogous
construction but  now with 
the $\sigma_3$ dependence constrained to be $Y^6 = -\sigma_3$. 
The second and fourth branes are also analogous but 
we impose $Y^5=-\sigma_2$. However if we try to consider the intersection
of the first and second branes in a similar manner 
then we must set $Y^4=\sigma_1$. As
a result the Bogomoln'yi equation \reef{slag} does not simplify to
a complex embedding condition. Instead we find the two-dimensional
Monge-Ampere equation
\be
\partial^2_2 U \partial^2_3 U 
- (\partial_2\partial_3 U)^2= 1\ ,
\label{MA}
\ee
where as before we set $Y^{i+3} =\partial_{i}F$ and take 
$F = {1\over2}\sigma_1^2 + U(\sigma_2,\sigma_3)$. Analogous equations
arise for the other intersections involving the first brane.

From the symmetry of the original problem it is clear that there should
be no physical difference between smooth intersections of the first brane
with the others and the complex intersections constructed above. Indeed
let us  change coordinates to\footnote{I thank N. Constable and W. Taylor
for suggesting this change of coordinates.} 
\be
\tilde w = {\rm Im}(Z)-i{\rm Re}(w) \ , \quad
\tilde Z = -{\rm Re}(Z) - i{\rm Im}(w)\ .
\label{newstatic}
\ee 
Note that now neither $\tilde w$ or
$\tilde Z$ are simply worldvolume coordinates or embedding coordinates.
After a little calculation however one can show that the condition
$\bar{\tilde{\partial}} \tilde Z=0$ becomes
\be
\partial Z = \bar \partial  \bar{ Z}\ ,\quad
|\partial  Z|^2-|\bar{\partial} Z|^2 =1\ .
\label{MAtwo}
\ee
The first condition implies that 
$Z = 2 \bar{\partial} U$ where $U$ is a real function. We
now find that the second condition in \reef{MAtwo} 
is precisely the Monge-Ampere equation \reef{MA}. 

To illustrate this point one can see that taking $\tilde Z(\tilde w)$ 
to have the form \reef{Zsol} 
(and setting $w_0=0$, $\alpha^2$ real and  $\lambda^2 =1$ for simplicity) 
produces the solution
\be
Z =\pm \sqrt{w^2 + \alpha^2{w\over \bar {w}}}\ ,
\label{nonhol}
\ee
under the change in static gauge given in \reef{newstatic}.
In this case the corresponding 
function $U$ only depends on $\sigma_2^2+\sigma_3^2$
and can be readily seen to solve \reef{MA}.
Note that near $w=0$, $Z \sim |\alpha| 
\sqrt{w / \bar {w}}$
is non-vanishing although 
$\partial Z$ and $\bar \partial Z$ diverge, {\it i.e.} at 
$w=0$ there is a circle of radius $|\alpha|$. 
This is a failure of static gauge and is easily resolved by
reflecting the solution across this circle and corresponds to choosing
the opposite sign for the embedding in \reef{nonhol} (recall that in the
holomorphic case this corresponds to going onto the second sheet of the
Riemann surface). A sketch of this intersection 
is therefore identical to figure 1 but rotated by
$\pi/2$.

Hence solutions to the Monge-Ampere equation can be found by
taking any holomorphic embedding $\tilde Z(\tilde w)$. 
We note that
the converse is not true. For example 
one can check that starting with
$\tilde Z(\tilde w,\bar{\tilde w})$ of the form \reef{nonhol} 
also leads to a 
solution $Z(w,\bar w)$ of \reef{MA} 
(this time $U$ only depends on $\sigma_2^2-\sigma_3^2$). 
Although again one can find coordinates in which this embedding is
holomorphic. Therefore the non-linearity in the 
Monge-Ampere equation merely reflects an awkward choice of static gauge.

\subsection{Three D3-branes Intersecting Over a Point}

In the previous section we explicitly constructed solutions to
the embedding equation \reef{slag} that described the resolution of any 
pairs of branes in \reef{figtwo} in single smooth complex surface.
Next we would like to obtain solutions to the embedding equation
\reef{slag} which depend non-trivially on all three coordinates 
and describe  smooth deformations
of three intersecting branes. 

\subsubsection{$SO(3)$ symmetry}

We consider the simplest possibility of a spherically symmetric
ansatz, namely $Y^{i+3} = \partial_i F(r)$
where $r=\sqrt{\sigma_1^2+\sigma_2^2+\sigma_3^2}$. Substituting this
into \reef{slag} leads to a second order differential equation which can
be integrated once to give
\be
G^3+3G^2-3G-1= {u^3\over r^3}\ ,
\label{slagsol}
\ee
where $G = F'/r$ and $u$ is an arbitrary constant. A plot of the cubic
equation is given in figure 2.
The embedding is then found by $Y^{i+3}=\sigma^i G(r)$.
Thus for a generic value of $r$ we find three solutions which can
be interpreted as three D3-branes. As $r\rightarrow\pm\infty$
the solutions are $G=1,-2-\sqrt{3},-2+\sqrt{3}$. This corresponds
to three flat D3-branes. In particular $G=1$ describes a D3-brane
in the $X^1,X^2,X^3$-plane while the other two solutions correspond
to two D3-branes in the planes $(X^4,X^5,X^6)=\pm\sqrt{3}(X^1,X^2,X^3)$.
This is precisely the intersection considered in \cite{gt}. 
Just as in the Monge-Ampere example, $r=0$ is not a
singularity but an artifact of static gauge. Instead we see
that, as $r\rightarrow 0$, $Y^{i+3} = u\sigma^i/r$, {\it i.e.} the coordinates
lie on a 2-sphere. We can construct a smooth manifold by attaching
another copy of the solution, where $r$ is negative and gluing it to
this 2-sphere.
Our resident artist's impression of this curve is given
in figure 3, but some explanation is required. First we start at
on the D3-brane corresponding to $G=1$ at $r=+\infty$.  As 
$r$ decreases (without loss of generality we assume that $u>0$) but 
is still positive we move
left on figure 3 until $r$ goes to zero. However we can smoothly pass
here to $r<0$ and we continue on the left of figure two until we
reach $G=-2-\sqrt{3}$ as $r\rightarrow -\infty$. This is plotted as the top
part of the curve in figure 3. Next we construct another branch of
the solution by starting at $G=-2-\sqrt{3}$ and $r=\infty$ and then
moving further right on
figure 2 passing through a minimum value of $r$ ({\it i.e.} a
maximum value of $G$) and finally out to $G=-2+\sqrt{3}$ as 
$r\rightarrow\infty$ again. 
This is plotted as part of the curve on the right and 
below the horizontal in figure 3. Finally a third branch is constructed
if we continue on  moving
right on figure two from $G=-2+\sqrt{3}$ as $r\rightarrow -\infty$, 
through a maximum value of $|r|$
({\it i.e.} a minimum value of $G$)
again until we get to $G=1$ at $r=-\infty$. This is plotted as the part of the
curve on the left and below the horizontal axis in figure three.
The rest of the curve arises as a result of the spherical symmetry.
In particular sections of the curve which are refected into each other
through the origin are in fact connected by the $SO(3)$ symmetry in 
the full six dimensions.

Thus the curve plotted in figure 3 
apparently smoothly interpolates between 
all three pairs of branes which are are $2\pi/3$ angles with each other.
However one can see from figure 3 that each of the three
D3-branes is  deformed in two directions. Clearly a single D3-brane can
only choose one such way to bend. This can be overcome by considering
two of each of the D3-branes, which are asymptotically parallel and 
coincident. The full curve
then represents the blow up of six D3-branes into a smooth 
Special Lagrangian.

\begin {figure}
\centerline{\epsfig{file=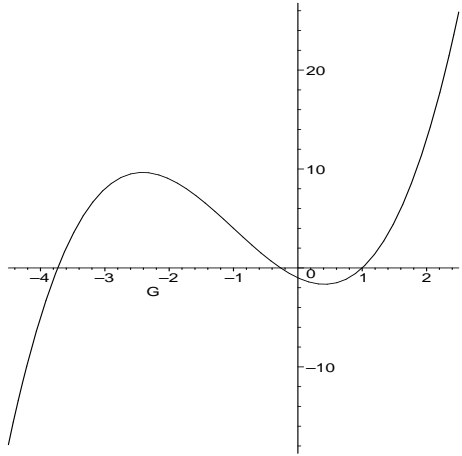,height=6cm,width=6cm}}
\caption{$u^3/r^3$ as a function of $G$}
\end{figure}

\begin{figure}
\centerline{\epsfig{file=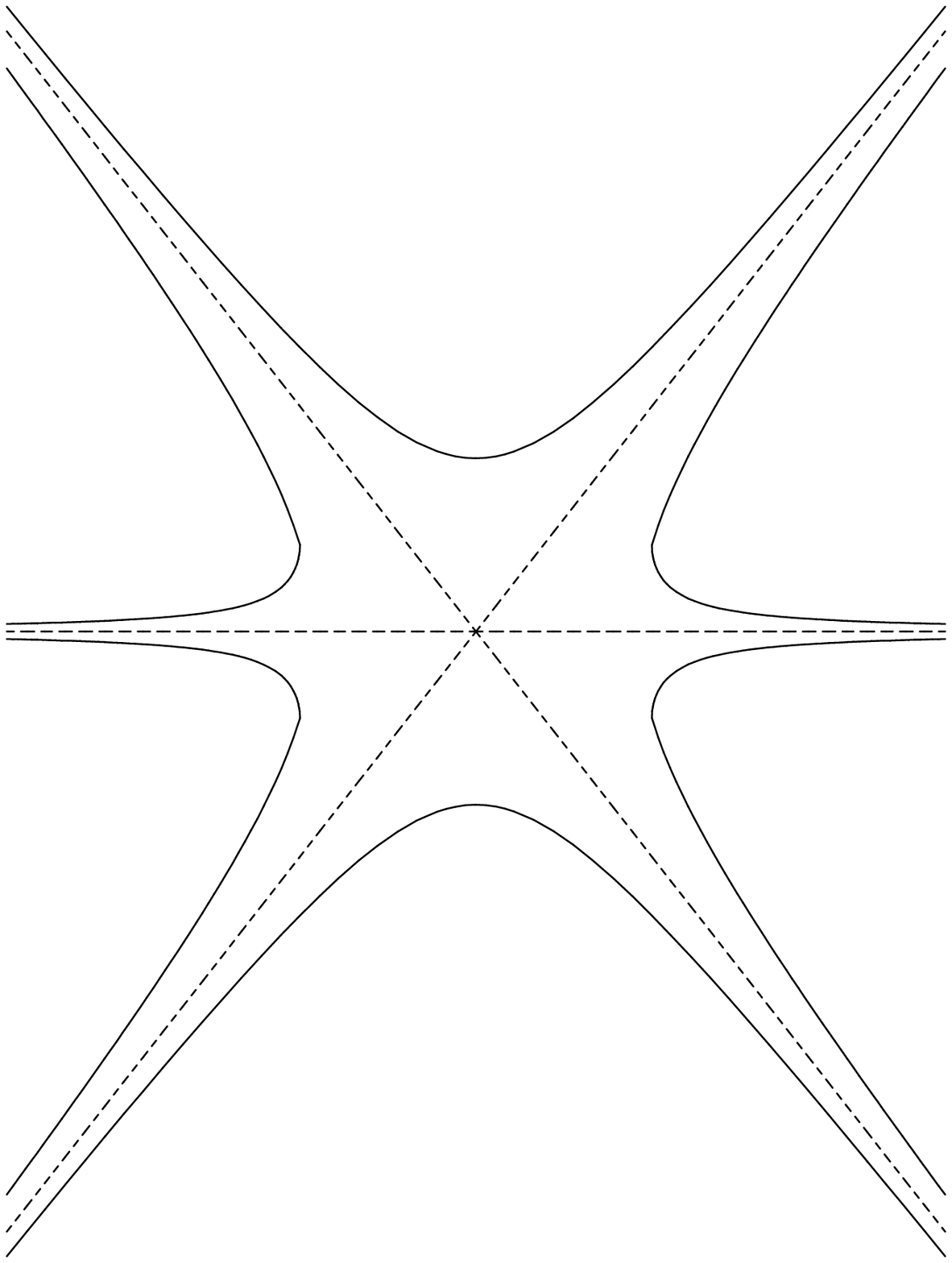,height=6cm,width=5cm}}
\caption{6 D3-branes embedded as a special Lagrangian}
\end{figure}

\begin {figure}
\centerline{\epsfig{file=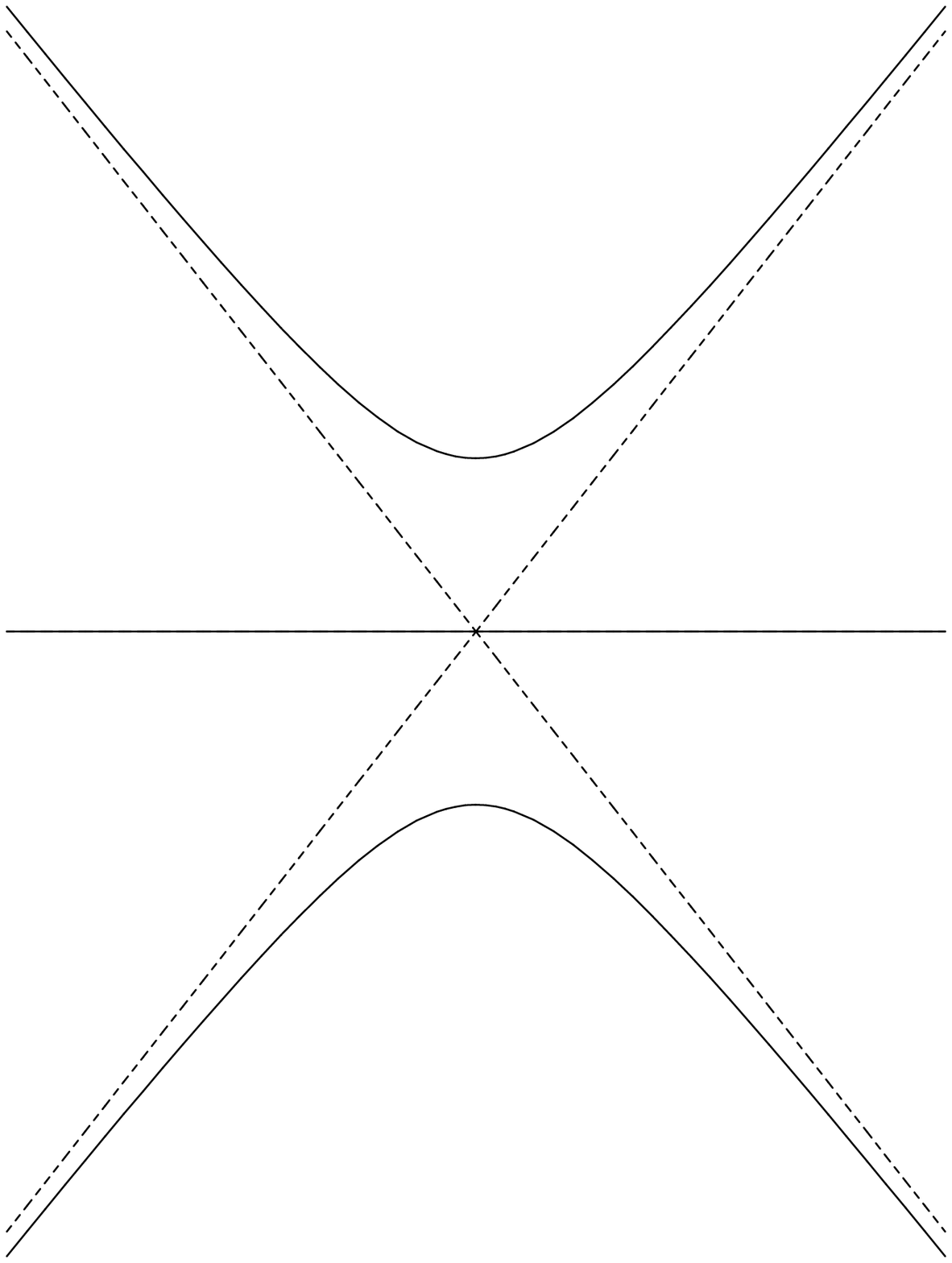,height=6cm,width=5cm}}
\caption{3 D3-branes embedded as a special Lagrangian}
\end{figure}

\begin {figure}
\centerline{
\epsfig{file=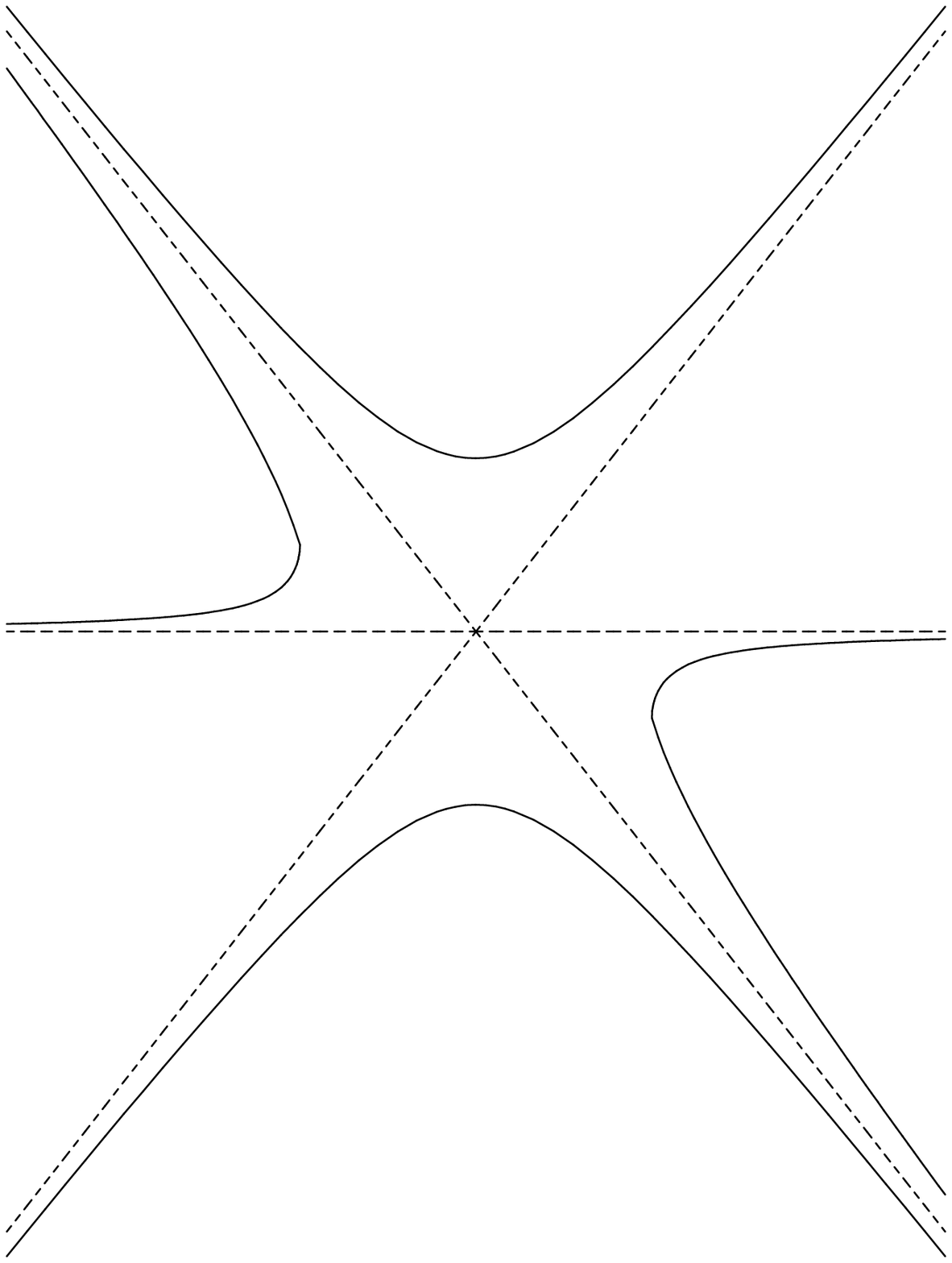,height=6cm,width=5cm}}
\caption{4 D3-branes embedded as a special Lagrangian}
\end{figure}

If we recall that
$X^i = (Y^i+Y^{i+3})/\sqrt{2}$ 
and $X^{i+3} = (Y^i-Y^{i+3})/\sqrt{2}$ then 
we can define
\be
\phi^i_1 = \sqrt{\frac{1}{3}} \frac{1-G}{1+G}X^i\ ,\quad 
\phi^i_2 = \sqrt{{3}}\frac{1+G}{1-G}X^{i+3}\ .
\label{GTcoords}
\ee
Now \reef{slagsol}  becomes
\be
\vec{\phi}_1\cdot \vec{\phi}_2 
= -{\rm sgn}(G^2-1)|\vec{\phi}_1||\vec{\phi}_2|\ ,
\quad 
|\vec{\phi}_1|(3|\vec{\phi}_1|^2-|\vec{\phi_2}|^2) =\rho\ ,
\label{gtcurve}
\ee
where $\rho = u^3/\sqrt{6}$. For $|G| >1$ this is the 
equation for the curve first discussed in  \cite{HL} and more recently
in \cite{gt}. In particular the construction of \cite{gt} consists of
keeping only 
one section of the curve, which smooths out the intersection between  two
D3-branes, while the third brane remains
flat and shoots through the hole created by the others. The resulting
embedding is sketched in figure 4. Another constuction of \cite{gt} 
is to simply delete
the flat brane that shoots through the middle, {\it i.e.} just to consider
the intersection of two D3-branes.

In 
a sense the region $|G|\ge 1$ contains one and a half
copies of the curve discussed in \cite{gt} since, in addition to a patch which
interpolates between one D3-brane with $G=1$ and another with 
$G=-2-\sqrt{3}$, the region $|G|\ge 1$ also includes a portion
of the curve that approaches the brane at $G=-2-\sqrt{3}$ 
and then, apparently,  abruptly ends
at $G=-1$. To resolve this we may extend to all values of $G$ to get the
embedding sketched in figure 3. In effect
this contains three copies of the curve in \cite{gt},
whereby all three branes are smoothed out. Alternatively we may
only extend as far as $G=-2+\sqrt{3}$ which corresponds to 
keeping only two sections of the curve in figure 3
and again all three D3-branes  are smoothed out. This is sketched in
figure 5.  However here again the middle D3-brane is deformed in
both directions and thus must be replaced by a pair of  
asymptotically parrallel and coincident
D3-branes. Hence it really represents the embedding of four seperate
D3-branes.

\subsubsection{$U(1)$ symmetry}

Next we would like to discuss intersections which are invariant only
under a $U(1)$ symmetry. In particular 
recently
special Lagrangian surfaces in ${\bf C}^3$ which are invariant under
$(Z^1,Z^2,Z^3) \rightarrow (Z^1,e^{i\theta}Z^2,e^{-i\theta}Z^3)$ 
have been  discussed in relation to mirror symmetry \cite{Joyce}.

If we consider the coordinates that we introduced in \reef{newstatic}
then, under this action, 
\be
\tilde w \rightarrow e^{-i\theta}\tilde w\ ,\quad
\tilde Z \rightarrow e^{-i\theta}\tilde Z\ .
\label{symmetry}
\ee
Therefore our task is to generalise the embeddings
in section 3.1
to include a non-trivial dependence on $\sigma_1$
and a non-holomorphic dependence on $\tilde w$
while respecting the symmetry \reef{symmetry}. Therefore the most
general ansatz is to write
$\tilde Z = \tilde w A({\tilde w\bar{\tilde w} },\sigma_1)$, where $A$ is
an arbitrary function. 
Examining the infinitessimal version of 
the transformation \reef{symmetry}, 
and using the fact that $\partial_2Y^6=\partial_3 Y^5$,  
we learn
that $|\tilde Z|^2-|\tilde w|^2 = 2a(\sigma_1)$. 
Hence we find 
\be\label{Aform}
\tilde Z = \pm\tilde w\sqrt{1+{2a\over {\tilde w\bar{\tilde w} }}} 
e^{i\theta}\ ,
\ee 
where $\theta$ is real and depends on $\sigma_1$ and $\tilde w\bar{\tilde w}$.
Note that the dependence of $\tilde Z$  on $\tilde w$ at a fixed value
of $\sigma_1$  is similar to the  
embedding \reef{nonhol}. 
One simple ansatz for a solution is to take $\theta$ and $a$ to be constant
and $Y^4$ linear in $\sigma_1$. This gives 
example 5.2 in \cite{Joyce} (with $\gamma=0$) and 
is essentially the holomorphic embedding \reef{Zsol} written in 
peculiar coordinates.
Taking the limit  $|\tilde w|\to\infty$ and solving \reef{slag} we find 
\be
M \to\left(\matrix{
{1+{\rm tan}(\theta)\over 1-{\rm tan}(\theta)}&0&0\cr
0&\pm{{\rm cos}(\theta)\over {\rm sin}(\theta)}&\pm{1\over {\rm sin}(\theta)}\cr
0&\pm{1\over {\rm sin}(\theta)}&\pm{{\rm cos}(\theta)\over {\rm sin}(\theta)}\cr
}\right)\ .
\ee
This corresponds to a smooth intersection of two 
D3-branes which asymptotically lie in the planes defined by
\be\label{plane}
X^4= \frac{1}{{\rm cos}(\theta)}X^2\pm {\rm tan}(\theta)X^3\ ,\quad 
X^5= \frac{1}{{\rm cos}(\theta)}X^3\pm {\rm tan}(\theta)X^2,\quad 
X^6 ={\rm tan}(\theta)X^1\ .
\ee

Less trivial solutions are more difficult to 
construct. However an existance proof has been given in \cite{Joyce} 
(in addition we learn that $a$ is always  constant).
Nevertheless the fact that these solutions take the form 
\reef{Aform} allows us to make some
remarks about the D3-brane interpretation.  
Just as above the apparent divergence at $\tilde w=0$
is an artifact of static gauge which is cured by adding a copy of the curve
corresponding to the opposite choice of sign in \reef{Aform}. 
Assuming that $\theta$ is independent of $|\tilde w|$ as 
$\tilde w \to \infty$ we see that for a fixed value of $\sigma_1$ 
there are two 
flat D3-branes which asymptotically take the form \reef{plane}. 
Thus these special Lagrangians
involve two intersecting D3-branes
which are in a sense twisted and tilted along $\sigma_1$.

\subsection{Moduli and  Superpotentials}

The existence of smooth solutions that represent intersecting branes
is a signal that there is a Higg's branch in the gauge theory living
on the intersection.
Atiyah and Witten \cite{aw} have argued  that, at least for
the case of three D-branes intersecting at $2\pi/3$ angles, there is a
superpotential
\be
W =  \Phi_{12}\Phi_{23}\Phi_{31}\ ,
\label{W}
\ee
where, for example,  $\Phi_{12}$ is the chiral multiplet
associated to the open strings that stretch from the first to the second
D3-brane. 
Thus the only flat directions along the Higg's branch are when one and only one
scalar field is non-vanishing, in which case the other
two scalar fields become massive. 

The existence of such a potential implies that the space of classical
smooth solutions to the calibration equation \reef{slag} has several branches.
Choosing to give a vev to a  
particular multiplet, say $\Phi_{12}$, is naturally
interpreted as smoothing-out the intersection of the first and 
second branes. However, 
due to the potential, once this is done 
the other smoothing-out modes are massive. A result of this interpretation
is that  there should be no smooth solution that interpolates
between all three D3-branes, but smooth solutions between any two pairs
of branes should exist. We note that the superpotential \reef{W} is supposed
to arise as a non-perturbative disk instanton effect \cite{aw} and 
therefore it may  seem unlikely that its effects 
can be visible in the classical description of special Lagaranian
surfaces presented here. 
However the disk instanton in question has zero-size and 
the superpotential  accordingly is not surpressed by any factor of $\alpha'$.
Thus one may expect that the classical geometry is sensitive to the 
superpotential.
In any case we will see that the solutions we found above are naturally
classified in terms of the flat directions of a non-Abelian superpotential 
on the moduli space.

For completeness we first consider the case of two 
D3-branes over a line.
Since these configurations 
in fact preserve eight supercharges a superpotential
is forbidden. The moduli space is then simply that of complex
curves with the relevant boundary conditions. 
The holomorphic condition $\bar\partial Z=0$ implies that the effective
Lagrangian \reef{modact} is simply 
\be
{\cal L}_{eff} = \int d^2 w {\partial Z\over \partial u^a}{\partial \bar Z\over\partial  u^b}
\partial^\mu u^a \partial_\mu u^b\ .
\label{bhbh}
\ee
Note that since 
two D3-branes intersect over
a line, it is natural to allow the moduli  $u^a$ to depend on this
spatial direction as well as time. Hence ${\cal L}_{eff}$ is a 
two-dimensional Lagrangian and we have accordingly introduced 
$\mu=0,1$ to label these directions. 
It follows from our discussion in section 2 
that \reef{bhbh} diverges and hence 
none of the moduli in \reef{Zsol} are normalisable. 
However by introducing additional
branes in various ways one can find examples where 
some of the moduli can be made normalisable and the 
$\sigma$-model produces a non-trivial
effective action \cite{HLW,dBHOO,LW2d}. 

Returning to  the case of D3-branes at $2\pi/3$ angles 
and the solution \reef{slagsol} we find that, 
in contrast to perpendicularly intersecting branes, 
the modulus $u$ is normalisable. Furthermore
one can show that \reef{modact} becomes 
\be
{\cal L}_{eff} = \kappa u^3{\dot u^2} 
\ ,
\ee
where $\kappa$ is a finite, positive constant.
We noted above that \reef{slag} and hence the associated modulus $u$  
led to three interpretations, although this required having parallel
pairs of D3-branes rather than three single D3-branes with Abelian gauge
fields.

We can  understand the appearances of these interpretation from 
the gauge theory potential \reef{W} 
as follows. If we have $N$, $N'$ and $N''$ of each
D3-brane then the moduli coming from stretched strings 
are in bifundamental representations of 
the $U(N)\times U(N')\times U(N'')$ 
gauge group of the D3-branes. In particular the analogue of \reef{W} is
\be
W = \Phi^{ab'}_{12}\Phi^{b'c''}_{23}\Phi^{c''a}_{31}\ ,
\label{Wtwo}
\ee
where $a = 1,..,N$, $b'=1,...,N'$ and $c''=1,...,N''$.
The supersymmetric vacuum moduli space ({\it i.e.} extrema of $W$)
is now much larger than the Abelian case and satisfies
\be
\Phi^{ab'}_{12}\Phi^{b'c''}_{23}=\Phi^{b'c''}_{23} \Phi^{c''a}_{31}=
\Phi^{c''a}_{31}\Phi^{ab'}_{12}=0\ .
\label{vacuum}
\ee 
If we write $\Phi_{12} = V_{12}\otimes V'_{12}$, 
$\Phi_{23} = V_{23}'\otimes V''_{23}$ and $\Phi_{31} = V_{31}''\otimes V_{31}$
then \reef{vacuum} becomes
\be
(V'_{12} \cdot V'_{23})  V_{12}  \otimes V''_{23} =
(V''_{23}\cdot V''_{31}) V'_{23} \otimes V_{31}   =
(V_{12}  \cdot V_{23})   V''_{31}\otimes V'_{12}  = 0
\ .
\label{vacuum2}
\ee

Therefore we can identify three types of branches of the supersymmetric vacuum 
moduli space. Firstly the branches which existed in the
Abelian case are still present. Here only one scalar field, say $\Phi_{12}$,
is non-vanishing. Up to a gauge transformation there are three, one-dimensional
branches of these vacua. Secondly  there is a symmetric 
branch where none of the
scalar fields vanish. In these cases  
$V_{12}\cdot V_{23}=V'_{12}\cdot V'_{23}=V''_{23}\cdot V''_{31}=0$.
Up to 
gauge transformations there is an $(N-1)(N'-1)(N''-1)$-dimensional
branch of such vacua. 
Lastly there will also be three mixed branches where
only one scalar field vanishes. For example if $\Phi_{31}=0$ then we
find $V'_{12}\cdot V'_{23}=0$ and hence, up to gauge transformations,
there is an  $(N'-1)$-dimensional branch of the vacuum moduli space.
Similarly there are also $(N''-1)$-dimensional and  $(N-1)$-dimensional 
branches for
$\Phi_{12}=0$ and $\Phi_{23}=0$ respectively. Note that the existance
of one of these mixed branches  requires that only
one of gauge groups is non-Abelian.

For the case at hand where
$N=N'=N''=2$ all of these vacuum branches are one-dimensional and we
can identify them with  the explicit curves found in section 3.2.
In particular the Abelian branches, where
only one scalar is choosen to be non-vanishing, 
correspond to the curve studied in \cite{gt},
where one pair of D3-branes is blown up and the third shoots through
the middle (figure 4). 
The three branches are obviously associated with the
three choices of which branes to smooth out. We also considered a curve
(figure 3)
which smooths out all three D3-branes but noted that this required having
two copies of each D3-brane. This corresponds to the symmetric vacuum branch.
Finally we found curves (figure 5) where all three
D3-branes were smoothed out, but not in a symmetric manner.
Clearly the three possible choices for which brane is in the middle 
can be identified with with the three mixed branches of the vacuum moduli 
space.

Finally we note that the pairs of parallel D3-branes coincide at infinity. 
It is natural then to ask if there is an associated solution
where they are separated. To answer this we note that 
separating a particular pair of D3-branes corresponds
to giving a non-zero vev to one of the adjoint scalar fields. Since the
$\Phi_{\alpha\beta}$ modes are in bifundamental representations 
of the gauge group they
will become massive via a Higg's effect. Therefore in such a 
vacuum the  relevant $\Phi_{\alpha\beta}$ scalars must vanish. Hence 
there should be no moduli correpsonding to seperating the branes.

\section{Relation to non-compact $G_2$ manifolds}

The discussion above focused on D3-branes. However nothing substantial
changes if we instead consider
D6-branes which are also extended along $X^7,X^8,X^9$. 
As first pointed out in \cite{townsend} (see also
\cite{Han,Duff}), upon
lifting these D6-branes to M-theory we find a solution which is
pure geometry ({\it i.e.} the four-form field strength vanishes).
The fact that four-dimensional $N=1$ supersymmetry is preserved along
the $X^0,X^7,X^8,X^9$ directions further implies that the remaining
seven-dimensional geometry $M$ has $G_2$ holonomy.

From the point of view of special Lagrangian surfaces the three types
of $SO(3)$ invariant 
solutions that we have described seem rather trivially related. 
In particular they simply correspond to various truncations of the
curve plotted in figure 3.
However upon lifting to M-theory these curves are associated to
different $G_2$ manifolds. To see this we can determine the topology
of the $M$ using the elegant formula derived in \cite{gs}
\be
h_2(M) = h_0(L)-1\ ,\quad h_{\alpha+2}(M) = h_\alpha(L) \ ,\  
\alpha=0,1,2,3\ ,
\ee
where $L$ is the special Lagrangian. The three special Lagrangian
embeddings $L$ that we discussed in section 3.2.1 have 
$(h_0(L),h_2(L))=(3,3),(2,1)$ and $(2,2)$ for figures 3,4 and 5
respectively, with all other Betti numbers vanishing. 
Therefore, upon lifting to M-theory, they correspond
to $G_2$ manifolds $M$ such that  
$(h_0(M),h_2(M),h_4(M))=(1,2,3),(1,1,1)$ and $(1,1,2)$
respectively,  with all other Betti numbers vanishing. 
The second case
is known to correspond to an ${\bf R}^3$ bundle over ${\bf CP}^2$ 
\cite{aw,gt}. The other manifolds have $h_4>1$ and thus cannot be
 ${\bf R}^3$ bundles over a smooth four-dimensional surface, although
they might be related to resolutions of the orbifold 
constructions in \cite{bw}.
Furthermore the existance of the large vacuum moduli space
discussed in section 3.3 for bigger gauge groups, {\it i.e.} many parallel 
D3-branes, suggests that there
is a wealth of additional $G_2$ manifolds related to the special 
Lagrangian curve \reef{slag}. However it is less clear what the geometrical
picture of these additional deformations is.

The $U(1)$ invariant solutions we discussed also lift to $G_2$ manifolds.
These have $(h_0(L),h_2(L))=(1,1)$ and hence lift to manifolds
with $(h_0(M),h_2(M),h_4(M))=(1,0,1)$. This suggests that their lift is
topologically an ${\bf R}^4$ bundle over ${\bf S}^4$.  Indeed a 
$G_2$ manifold with  this form was 
discussed in \cite{aw,gt} and is  the lift of
two intersecting D6-branes at $2\pi/3$ angles (corresponding
to the $SO(3)$ invariant curve in figure 4 with the 
flat D3-brane is deleted). Since the $U(1)$
invariant special Lagrangians describe two D6-branes lying in the
planes defined by \reef{plane} it is
natural to suppose  that 
that their associated $G_2$ manifolds can be continuously connected
to the ${\bf R}^4$ bundle over ${\bf S}^4$ of \cite{aw,gt}. 

\section{Conclusions}

In this paper we revisted the worldvolume analysis for intersecting
D-branes, including a discussion of the moduli space metric. We then
focused our discussion on D3-brane intersections  
which preserve four supercharges. In particular we found
a new embedding condition which treats all D3-branes on an equal footing.
We also discussed some solutions to this condition and identified their
moduli space with  the vacuum branches of a non-Abelian 
superpotential. Upon lifting to M-theory these solutions imply the
existance on new $G_2$ manifolds.

It would be of interest to find more explicit solutions to the special
Lagrangian embedding equation \reef{slag}. 
Since we have found 
all the solutions that are predicted by the superpotential \reef{W}
for the case of D3-branes at $2\pi/3$ angles, it seems natural to assume
that any new solutions will satisfy different boundary conditions. 
It would also be interesting to find solutions which represent 
all four intersecting D3-branes.
However we note that in the solition \reef{slagsol}
the three distinct D3-branes at infinity correspond to the three
solutions to a cubic equation. Since \reef{slag} is at most cubic
in the embedding coordinates it seems that solutions with more than
three D3-branes are
impossible. 
Finally it would be interesting to derive embedding
equations which to do not prefer one D-brane over the others for
configurations with less supersymmetry, such as associative, co-associative
and $spin(7)$ calibrations.

\bigskip
{\noindent \bf Acknowledgments}\\
I would like to thank B. Acharya, M. Douglas, S. Gukov, C. Hofman
and D. Tong for discussions and especially
N. Constable and  W. Taylor for their initial 
collaboration on this project. I would also like to thank the theoretical
physics groups at  MIT and Harvard for their hospitality 
while part of this work was completed.

\end{document}